\documentclass[12pt]{iopart}

\usepackage{graphicx}
\graphicspath{{./}}
\usepackage{textcomp}
\usepackage[english]{babel} 
\usepackage[utf8]{inputenc}
\usepackage[T1]{fontenc}
\usepackage{hyperref}
\usepackage[amssymb]{SIunits}
\usepackage{color}

\newcommand{\UGAAddress}{Univ. Grenoble Alpes, F-38000 Grenoble, France}
\newcommand{\CNRSAddress}{CNRS, Institut Néel, "Nanophysique et semiconducteurs" group, F-38000 Grenoble, France}
\newcommand{\CEAAddress}{CEA, INAC-SP2M, "Nanophysique et semiconducteurs" group, F-38000 Grenoble, France}

\begin{document}

\title[Coupling of plasmonic nanoantennas to nanowire quantum dots]{Deterministic radiative coupling between plasmonic nanoantennas and semiconducting nanowire quantum dots}

\author{Mathieu Jeannin$^{1,2}$, Pamela Rueda-Fonseca$^{1,3}$, Edith Bellet-Amalric$^{1,3}$, Kuntheak Kheng$^{1,3}$ and Gilles Nogues$^{1,2}$}
\address{$^1$ \UGAAddress}
\address{$^2$ \CNRSAddress}
\address{$^1$ \CEAAddress}

\date{\today}

\begin{abstract}
We report on the deterministic coupling between single semiconducting nanowire quantum
dots emitting in the visible and plasmonic Au nanoantennas. Both systems are
separately carefully characterized through microphotoluminescence and
cathodoluminescence. A two-step realignment process using cathodoluminescence
allows for electron beam lithography of Au antennas near individual nanowire
quantum dots with a precision of \unit{50}{\nano\meter}. A complete set of
optical properties are measured before and after antenna fabrication. They
evidence both an increase of the NW absorption, and an improvement of the quantum dot emission rate up to a factor two in presence of the antenna.
\end{abstract}
\noindent{\it Keywords\/}: semiconductor, nanowire, plasmonics, nanoantenna, nanofabrication, cathodoluminescence
\pacs{42.79.-e, 78.67.-n, 42.82.Fv}
\submitto{\NT}

\maketitle 

\section{Introduction}
Recent progress in semiconductor growth research now allows for fabrication of
nanowires (NWs) structures which are of great interest for their high
crystalline quality, strain free character and practical geometry. A key step
towards realization of nano-optical circuits and applications relies on the
coupling of single NWs with other structures, like photonic crystal cavities
\cite{Birowosuto2014} or plasmonic nanoantennas
\cite{Ozel2013,Casadei2014,Casadei2015,Ramezani2015} (NAs). To better use the
advantages and the versatility of NWs, it is now necessary to adapt previous
studies on self-assembled or colloidal quantum dots (QDs) to their NW counterparts. While
numerous results have been obtained in the coupling of self-assembled 
\cite{Curto2010,Pfeiffer2010,Nogues2013,Belacel2013,Kukushkin2014} and colloidal 
\cite{Curto2013,Hoang2016} QDs with
plasmonic nanostructures, all the
existing studies on control of the optical properties of nanowire quantum dots
(NWQDs) rely on a photonic approach, using the nanowire itself as an antenna
\cite{Claudon2009,Munsch2012,Cremel2014}. Following previous work in our group
\cite{Nogues2013} on droplet epitaxy QDs, we show how combining
cathodoluminescence (CL) with standard e-beam lithography technique allows
to fabricate at will plasmonic NAs in the vicinity of II-VI NWQDs emitting
around \unit{620}{\nano\meter}. Both systems are initially fully characterized
using CL for the antennas and micro-photoluminescence (\textmu PL) and time-resolved 
spectroscopy for the emitters. Previous studies have made use of morphological criteria to
detect nanoemitters, using scanning electron microscopy (SEM) or atomic force
microscopy \cite{Pfeiffer2010}. In contrast, our method has the advantage 
of allowing localization of embedded structures undetectable with the previous methods.

\section{Nanowire quantum dot properties}
\label{sec:NWproperties}
\begin{figure}
	\centering
	\includegraphics[trim = 0cm 0cm 0cm 0cm, clip = true, width=\linewidth]{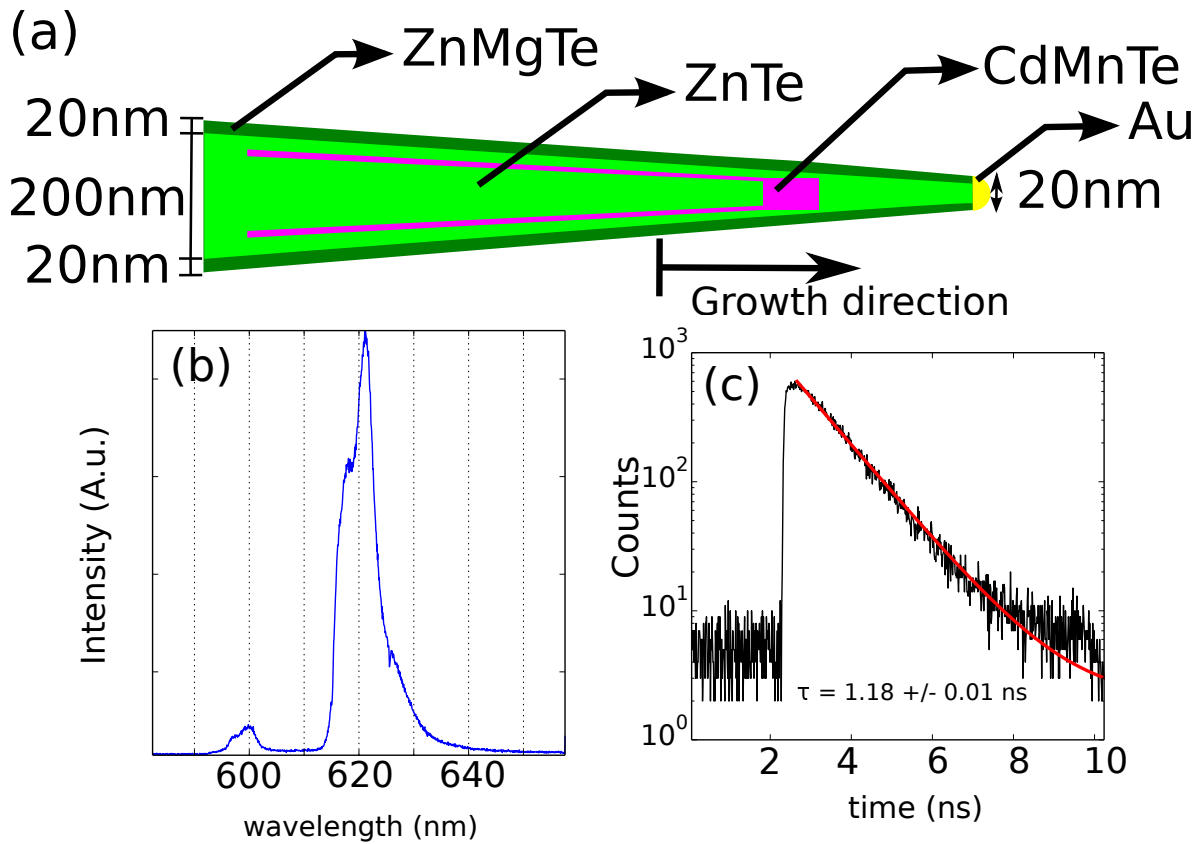}
	\caption{(a) Nanowires morphology: an Au droplet serves as a catalyst to grow a
		CdTe quantum dot inserted in a ZnTe conical nanowire. A lateral regrowth of
		ZnMgTe creates a shell around this core. (b) typical \textmu PL spectrum of a
		nanowire QD and (c) corresponding decay rate measurement, integrating a
		bandwidth of \unit{10}{\nano\meter} around \unit{622}{\nano\meter}. A
		monoexponential fit (red line) gives a lifetime $\tau
		=$ 1.18$\pm$\unit{0.01}{\nano\second}. }
	\label{fig:NW_presentation}
\end{figure}

Our nanoemitters are single (Cd,Mn)Te QDs (Mn fraction $\approx 5\%$) inserted
inside ZnTe/(Zn,Mg)Te core/shell nanowires [figure \ref{fig:NW_presentation}~(a)].
The NWs are grown by molecular beam epitaxy on a (111)B GaAs substrate covered
by a \unit{500}{\nano\meter} thick buffer layer of ZnTe \cite{Artioli2013,
	Rueda-Fonseca2014}. Dewetted gold droplets are used as catalysts, and
temperature variations allow to favor different growth mechanisms resulting in
the final heterostructure. The core growth is both longitudinal and lateral,
leading to the conical aspect of the nanowire, with a core diameter of
\unit{20}{\nano\meter} at the top to $\approx$\unit{200}{\nano\meter} at the
base, and a \unit{20}{\nano\meter} thick shell. The QD emission is measured by
confocal \textmu PL spectroscopy. It features a main emission peak around
\unit{620}{\nano\meter}, blue-shifted by \unit{150}{\nano\meter} compared to the
exciton transition in bulk CdTe \cite{Horodysky2005} due to the confinement and the 
strain induced by the surrounding shell \cite{Ferrand2014}. It is significantly
broadened (FWHM \unit{10}{\nano\meter}) by the presence of the magnetic Mn
atoms creating a fluctuating magnetic field inside the QD, randomly shifting the
exciton line in time by Zeeman effect \cite{stepanov:tel-00994939}  [figure
\ref{fig:NW_presentation}~(b)]. Amongst all the studied NWs, the central
emission wavelength ranges from 605 to \unit{630}{\nano\meter}.

NWs are detached from their growth substrate and dispersed onto a host substrate
for optical studies. The host is a Si substrate pre-patterned by optical
lithography and dry etching to fabricate coarse localization marks. It is
covered by a \unit{100}{\nano\meter} thick Au layer deposited by e-gun
evaporation and a \unit{250}{\nano\meter} thick Al$_2$O$_3$ spacing layer by
atomic layer deposition. The mirror and the spacer thickness are designed to
give constructive interferences between the light directly emitted from the QD and 
the light reflected by the mirror in order to 
maximize luminescence collection. NWs are finally dispersed by mechanical
contact between the host and growth substrates.

A first selection of NWs is performed using CL at \unit{5}{\kelvin} 
\cite{Nogues2013, Artioli2013}. The luminescence from the QD is spectrally
filtered and detected by an avalanche photodiode (APD). CL imaging of the
emission profile at the QD luminescence wavelength is obtained for a set of NWs. To
avoid spurious emission from exciton trapped in defects at the base of the NWs,
we conserve only structures presenting a well localized emission in the top half
of the conical NW and having the typical spectral structure of figure 
\ref{fig:NW_presentation}~(b). The selected QDs are then fully characterized by
\textmu PL spectroscopy at \unit{5}{\kelvin}. The NWs are excited at
\unit{447}{\nano\meter} by a frequency doubled, ps pulsed Ti:Sapphire.
Time-resolved measurements are performed at very low pumping power, 
well below saturation of the QD, with a spectral integration bandwidth of \unit{2}{\nano\meter} 
to ensure that we only detect the excitonic transition, 
resulting in a monoexponential time trace. They show decay times ranging from 0.2 to
\unit{1.18}{\nano\second} [figure \ref{fig:NW_presentation}~(c)]. The average lifetime ($\tau=$ 
\unit{0.68}{\nano\second}) is about five times longer than for self-assembled
CdTe/ZnTe QDs \cite{man2015}. In addition, the emission from the QD is linearly
polarized in a direction parallel to the NW axis in $70\%$ of the cases, and
orthogonal in all the others. The average degree of linear polarization is
$0.7\pm 0.2$. These optical properties can be explained by the very small energy
difference between the valence bands of CdTe and
ZnTe \cite{stepanov:tel-00994939}. The resulting nature of the ground hole state
strongly depends on additional energy shifts induced by strain or QD aspect
ratio \cite{Zielinski2013, Ferrand2014}. In our case, the orbital hole
wavefunction is probably poorly confined inside the QD, resulting in low
electron-hole wavefunction overlap and thus long radiative decay time. We note
that the measured decay rate $1/\tau = \gamma_r + \gamma_{nr}$, where $\gamma_r$
(resp. $\gamma_{nr}$) is the radiative (resp. non-radiative) decay rate, might
be limited by non-radiative processes. Furthermore, a large variability from dot
to dot strongly mixes the light- and heavy-hole bands, explaining the
polarization results. This variability is not detrimental to our study as we
compare a complete set of optical properties \emph{on the one and same NW before and
after antenna fabrication}.

\section{Nanoantenna characterization}
\label{sec:NACarac}
\begin{figure}
\centering
\includegraphics[trim = 0cm 0cm 0cm 0cm, clip = true, width=\linewidth]{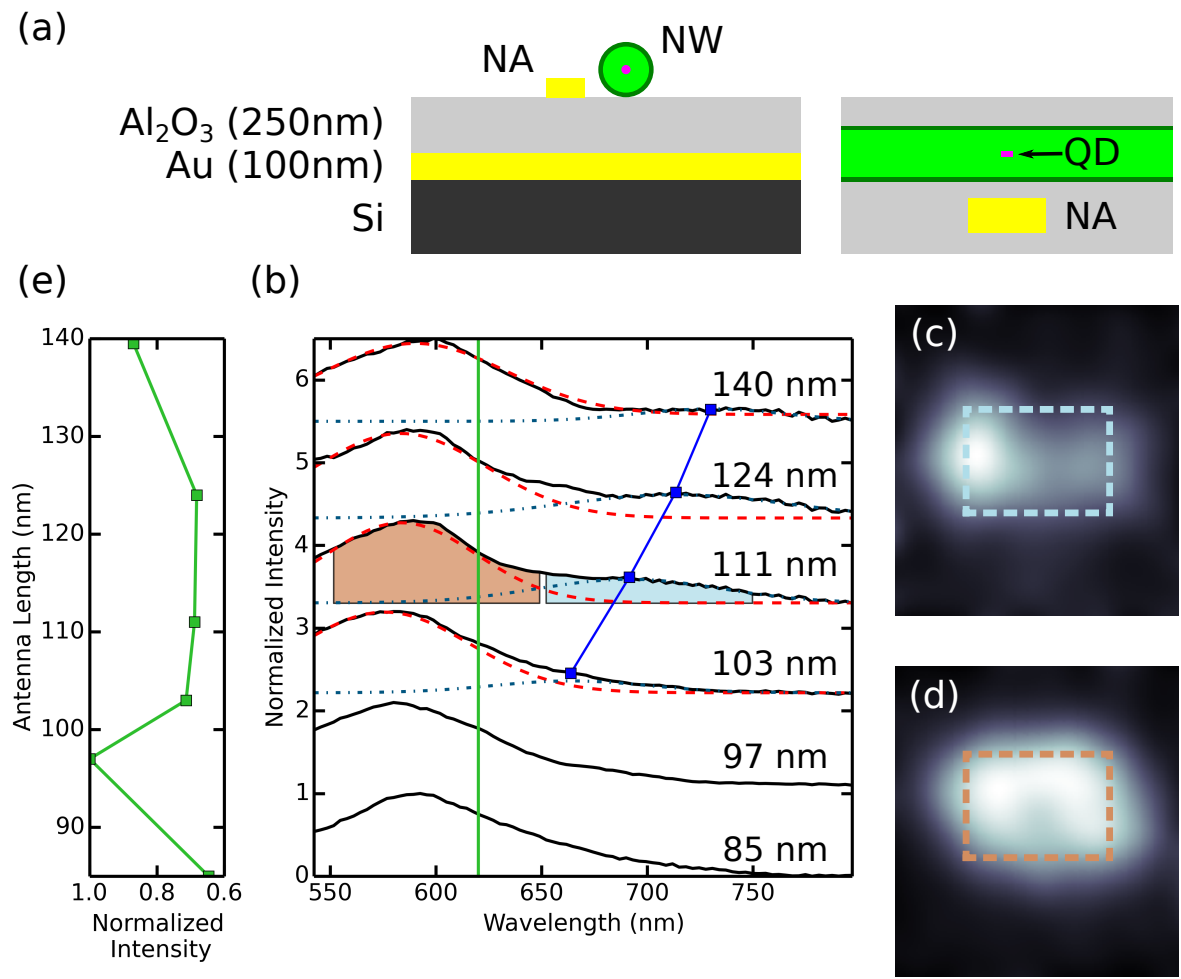}
\caption{(a) Schematic of the NA-NW geometry (left: side view, and right: top view), to scale.
	(b) Cathodoluminescence spectra of the rectangular nanoantennas, with Gaussian fits of the main 
	component (red dashes) and the dipolar mode (blue dashes) discussed in the text. 
	The green line corresponds to (e). The blue line with squares 
	indicates the fitted central wavelength of the
	dipolar resonance. The blue and brown shaded areas correspond to the 
	integrated bandwidth in (c) (blue) and (d) (brown) respectively.
	(c) and (d) LDOS imaging of the \unit{111}{\nano\meter} long
	antenna around \unit{700}{\nano\meter} and \unit{600}{\nano\meter} respectively,
	with \unit{100}{\nano\meter} spectral integration bandwidth. The blue and brown rectangles
	correspond to the antenna boundary. (e) Cut profile of the CL intensity along
	$\lambda_{QD}=$ \unit{620}{\nano\meter} [green solid line on (b)] showing an increase 
	of the measured LDOS for an antenna length of \unit{97}{\nano\meter} due to the 
	onset of the dipolar mode.}
\label{fig:Spectral_map}
\end{figure}

Au nanoantennas are fabricated on the host substrate, in a region empty of NWs.
They consist in single rectangles of \unit{70}{\nano\meter} fixed width, as sketched in 
figure \ref{fig:Spectral_map}~(a). 
Their length L varies from 85 to \unit{140}{\nano\meter}. They are fabricated by 
electron-beam lithography on a \unit{200}{\nano\meter} thick 
Poly(methyl methacrylate) (PMMA) bi-layer resist, using PMMA with 
molecular weights of \unit{50}{\kilo\dalton} and \unit{950}{\kilo\dalton} 
to create a mechanical mask with the resist.
\unit{35}{\nano\meter} of Au is deposited on the sample by e-gun evaporation.
N-methyl-2-pyrrolidone (NMP) lift-off with 80$^{\circ}$C heating is finally performed
to remove the resist.

The antennas are characterized using CL at room temperature 
\cite{GarciadeAbajo2010, Vesseur2007}. The CL spectrum of each antenna is obtained by
raster-scanning the electron beam over its surface. It presents a large peak at a fixed wavelength
around \unit{580}{\nano\meter} and a second resonance at lower energy. 
CL spectra for all the antenna are presented in figure \ref{fig:Spectral_map}~(b), revealing
that the second resonance red-shifts with increasing antenna length. We note
that the optimization of the Al$_2$O$_3$ spacer thickness in order to increase the
collected signal at \unit{620}{\nano\meter} results in destructive interferences
for the collected light in the 700 to \unit{800}{\nano\meter} wavelength range, 
decreasing the detection contrast in this spectral region.
Further information is obtained by imaging the plasmon local density of
states \cite{GarciadeAbajo2010} (LDOS). It is reconstructed by slowly scanning
the electron beam over the antenna and collecting the CL emission as a 
function of the beam position filtered in a \unit{100}{\nano\meter} spectral window 
around the resonance energy on an APD. This spectral 
integration bandwidth is chosen to maximize the collected signal from the antenna. For $L\geq$\unit{110}{\nano\meter}, it is 
smaller than the energy separation between two consecutive modes so that only a single 
resonance contributes to the LDOS image. 
The LDOS of the red-shifting resonance [figure \ref{fig:Spectral_map}~(c)] 
clearly displays two lobes characteristic of a longitudinal dipolar mode. 
In this regime the antenna sustains a single radiating mode whose
dispersion relation can be extracted from figure \ref{fig:Spectral_map}~(b)(blue squares). 
On the contrary, the LDOS at \unit{580}{\nano\meter} [figure \ref{fig:Spectral_map}~(d)] 
shows no precise spatial structure at any length. We have fabricated nanoantennas of different shapes, sizes and spacer thicknesses and observed that it is always present. The energy of its maximum only depends on the Al$_2$O$_3$ spacer thickness (see Supplementary Material \cite{Supplementary}). We attribute this peak to the scattering by the NA of the continuum of
surface plasmon polariton modes sustained by the Au/Al$_2$O$_3$/air multilayer system. 
LDOS evaluation based on numerical calculations of the surface plasmon polariton dispersion relation\cite{Davis2009} show that at this energy a lot of lossy
modes contribute to the signal. 
We note that we use here a \unit{250}{\nano\meter} thick spacer for which no significant coupling 
occurs between the antenna and the metal
film, which simply acts as a mirror to reflect emitted and scattered light
towards the collection objective. To determine the relevant dimensions for
enhancing emission processes at the QD wavelength, we plot a cut of figure 
\ref{fig:Spectral_map}~(b) at $\lambda_{QD}=$ \unit{620}{\nano\meter} [green solid 
line, figure \ref{fig:Spectral_map}~(e)]. It predicts an increase in scattering
efficiency for an antenna length around \unit{95}{\nano\meter} due to the contribution of the dipolar
mode. It also predicts an enhancement for lengths above \unit{140}{\nano\meter}.
This enhancement comes from the onset of a higher order mode theoretically predicted 
to appear at this length \cite{Davis2009,Filter2012}. We note however that for
the target coupled NW-NA system, the presence of the very high refractive index
nanowire (n$_{\mathrm{ZnTe}}\approx$~3) in the near-field of the antenna
significantly modifies its plasmonic properties. Hence a red-shift in the plasmon
dispersion relation curve is expected, and the enhancement peaks observed in
figure \ref{fig:Spectral_map}~(e) will occur for smaller antenna lengths.

\section{Hybrid NW-NA structures}

\subsection{Fabrication}
\begin{figure}
	\includegraphics[trim = 0cm 0cm 0cm 0cm, clip = true, width=\linewidth]{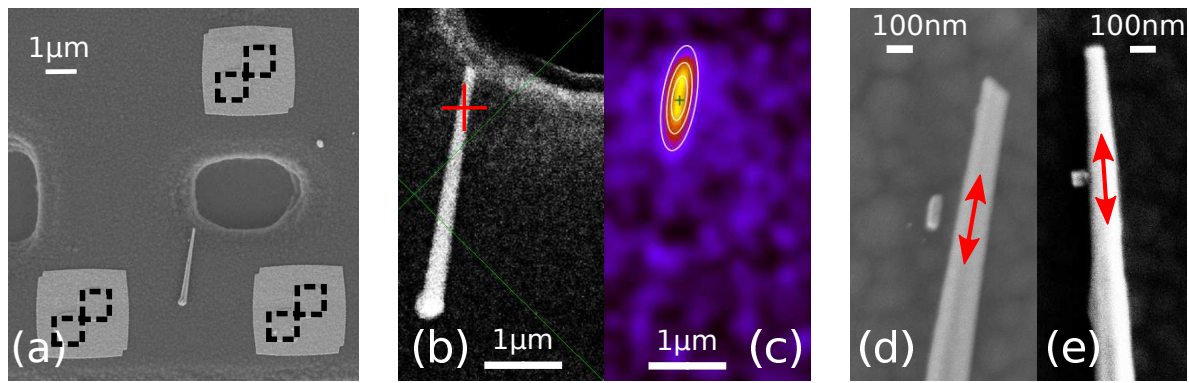}
	\caption{(a) Large field SEM image of the sample after antenna fabrication. The
		dashed lines indicate the position of the fine alignment marks which were
		exposed during the last alignment step and are hardly visible under the final
		Au layer. (b) Higher magnification SEM image of the NW after alignment with
		respect to the marks (c) Filtered CL intensity at
		$\lambda_{QD}=$ \unit{620}{\nano\meter} at the same position and magnification.
		The red cross in (b) indicates the center of the bi-dimensional Gaussian fit.
		The white lines in (c) are contour lines of the fit. (d) SEM image of the
		resulting coupled NW-NA system, along with (e) another NA-NW system. The NA-NW
		gap is \unit{60}{\nano\meter} in (d), while a point contact is achieved in (e).
		The red arrows indicates the polarization direction of the emitted light, which
		is ensured to be aligned with the antenna dipolar mode axis.}
	\label{fig:repos}
\end{figure}

Two examples of the target structure are shown in figures \ref{fig:repos}~(d)-(e).
Fine alignment marks are first fabricated around the chosen NWs using electron-beam 
lithography, metal evaporation and lift-off [figure \ref{fig:repos}~(a)]. 
We then record a SEM image [figure \ref{fig:repos}~(b)] 
together with the CL intensity image at $\lambda_{QD}=$ \unit{620}{\nano\meter} 
[figure \ref{fig:repos}~(c)].
Both images are acquired after aligning the microscope beam using the fine
alignment marks. Acquisition times are limited to a few seconds to limit
mechanical and electrostatic drifts. We fit the QD emission peak with a
bi-dimensional Gaussian profile and superimpose it to the corresponding SEM
image. It allows to precisely localize the QD inside the NW in the frame defined
by the fine alignment marks. NAs are then fabricated with the same process as
the previously characterized antennas. Before insulation of the resist, a final
alignment step is performed onto the fine marks under the resist. The antennas
have a fixed width $w=$ \unit{70}{\nano\meter} and a variable length L from $50$
to \unit{140}{\nano\meter}. This range is based on the previous CL results and
takes into account the shift due to the presence of the NW. The antenna sides of
length L are parallel to the measured polarization direction of the QD emission.
We aim at having an antenna to NW gap equal to zero. SEM images reveal an
average gap of \unit{12}{\nano\meter}$\pm$\unit{50}{\nano\meter}, thus some
antennas are on top of the NW. The final error of \unit{50}{\nano\meter} has
different sources. Alignment of the lithography setup has a typical error of
\unit{20}{\nano\meter} but is degraded in our case due to poor contrast of the
fine alignment marks image under the resist. We evaluate the thermal, mechanical
and electrostatic drifts error during CL at low temperature to
\unit{30}{\nano\meter}. We note that the NW core and shell are already thick
($\approx$\unit{50}{\nano\meter} at the QD position). The resulting QD to
antenna distance ranges from 50 to \unit{110}{\nano\meter}. This distance is 
always large enough to prevent luminescence quenching, and the variations due 
to the positioning error only have a moderate effect 
on the NA to QD coupling \cite{Anger2006}.

\subsection{Photoluminescence observations}
\begin{figure}
\centering
\includegraphics[trim = 0cm 0cm 0cm 0cm, clip = true, width=.9\linewidth]{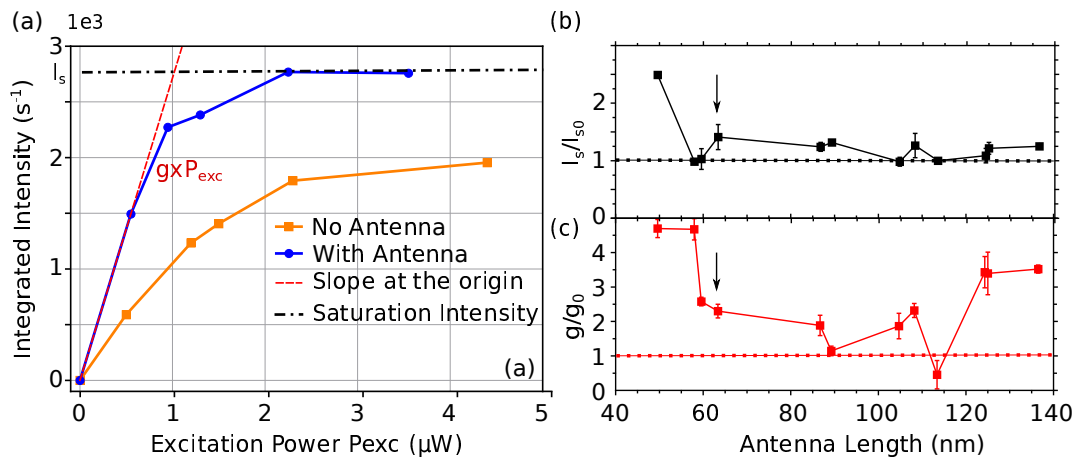}
\caption{(a) Integrated \textmu PL intensity for one NWQD as a function of
	excitation power before (orange square) and after (blue circle) nanoantenna
	fabrication. The NA is \unit{63}{\nano\meter} long. Each data set is fitted with
	two components, a constant saturation intensity $I_s$, and a linear function of 
	slope $g$ describing the QD emission at low 
	pumping power. The saturation 
	intensity is determined by the last data point. The slope at the origin 
	is determined between the first data point and the origin. 
	(b) Saturation intensity $I_s$ for different NA-NW system divided by the corresponding 
	value of the NW without antenna $I_{s0}$ as a function of antenna length L. 
	The error bars are given by the detection noise of the spectrometer.
	(c) Slope at the origin $g$ for different NA-NW system divided by the corresponding 
	value of the NW without antenna $g_0$ as a function of antenna length L. The error bars 
	are evaluated by comparing the slope at the origin and a linear regression 
	between the first two data points and the origin.
	The arrow in (b) and (c) corresponds to the NWQD presented in (a). 
	The dotted line in each figure indicates a factor of 1.}
\label{fig:PL_results}
\end{figure}

The coupled QD-NAs systems are characterized using \textmu PL spectroscopy with 
a pulsed excitation laser at a wavelength of \unit{447}{\nano\meter}.
Figure \ref{fig:PL_results}~(a) compares the integrated QD emitted intensity 
as a function of the exciting laser power $P_{exc}$ before and after
antenna fabrication for $L=$ \unit{63}{\nano\meter}. 
The PL intensity first increases with the excitation power and then 
saturates because of the complete occupation of the discrete excitonic 
state in the QD. For a same excitation power 
one clearly sees a much higher collected signal in presence of the NA. Similar
measurements on other QDs present on the substrate which have experienced the
same process except the final antenna fabrication show no change in their
properties. 
Previous studies revealed modification of the luminescence
collection by redirection of light by the antenna
\cite{Ramezani2015,Curto2010,Nogues2013}. We have carried out Fourier plane microscopy 
by imaging the back focal plane of the microscope objective in the \textmu
PL setup (see supplementary material \cite{Supplementary}) to observe the
radiation diagram of the NWQDs \cite{Grzela2012}, and we did not observed 
significant change in the radiation pattern after antenna fabrication. 
Clearly, the higher PL intensity we measure is not due to the redirection of light 
by the antennas. Additional information is provided by measurements of the 
exciton lifetime before ($\tau_0$) and after ($\tau_{NA}$) antenna fabrication 
under the same excitation conditions. We observe no significant change, 
with an average ratio $\tau_{NA}/\tau_{0} = 1.1 \pm 0.2$ 
(see Supplementary Materials \cite{Supplementary}).

To better analyse the effect of the antennas, we evaluate the saturation intensity $I_s$ at high
excitation power, and the slope at the origin $g$ of the power-dependent \textmu PL curve. 
The ratio of $I_s$ before and after antenna fabrication is
plotted as a function of antenna length in figure \ref{fig:PL_results}~(b), showing 
a net increase after NA fabrication.
As shown in figure \ref{fig:PL_results}~(c), we also observe a greater increase 
in the slope $g$ after NA fabrication. 

\subsection{Discussion}

Assuming that we only detect the excitonic transition and considering that the
nanowire excitation is done with a pulsed laser, the saturation intensity $I_s$ is
equal to $f\times \gamma_{r}/(\gamma_{r}+\gamma_{nr})\times\alpha_{coll}$, where
$f$ is the laser repetition rate, $Y=\gamma_{r}/(\gamma_{r}+\gamma_{nr})$ is the
quantum yield and $\alpha_{coll}$ is the fraction of the emitted power collected
by the microscope objective. As we measure no change in the radiation diagram after NA
fabrication, it is reasonable to assume that $\alpha_{coll}$ is unperturbed.
As a consequence the enhancement observed in figure \ref{fig:PL_results}~(b) is
directly related to an increase in the quantum yield $Y$. This enhancement is found to be moderate 
for antennas with a length greater than \unit{60}{\nano\meter} but stronger for the smallest 
antenna, up to a factor $2.5$. As expected from section \ref{sec:NACarac}, it occurs 
for shorter antennas than the determined resonant length from figure \ref{fig:Spectral_map}~(e). 
Moreover since the QD response at saturation is independent of the excitation power, the change of $I_s$ after antenna fabrication 
is not due to an increase of the absorption inside the NW.

Nevertheless, CL spectra of the 
NAs at $\lambda_{exc} = $\unit{447}{\nano\meter} show that there is still a measurable LDOS 
at this energy that can modify the absorption of the excitation laser inside the NW 
(see Supplementary Material \cite{Supplementary}). The main contribution to this LDOS comes from the fixed resonance observed in Figure \ref*{fig:Spectral_map} and does not depend on the antenna length.
The change of the absorption can be retrieved by analyzing the slope at the origin $g$.
In the limit of low excitation regime, $g$ is expected to be proportional to 
$Y\times \alpha_{abs} \times \alpha_{coll}$, 
where $\alpha_{abs}$ is the fraction of power absorbed in the
NW. 
As $\alpha_{coll}$ remains unchanged after NA fabrication, the ratio $(g/g_0)/(I_s/I_{s0})$  
directly gives the change in $\alpha_{abs}$. 
The comparison of figures \ref{fig:PL_results}~(b) and (c) shows that the absorption in the NW is 
enhanced by a factor 2.2$\pm$1.1 over the whole antenna length range.
Further illustration of the respective changes in $I_s$ and $g$ are presented in the 
Supplementary Materials \cite{Supplementary} with two other power-dependent \textmu PL 
curves. Our main conclusion is that, for the shorter antenna lengths, enhanced NW absorption and improvement of the quantum yield  contribute equally at $\sim$50\% to the four-fold increase of $g$ that we observe.

Finally, lifetime measurements show that there is no change of the total decay rate 
$\gamma_{r}+\gamma_{nr}$. This is an indication that the total decay rate is dominated 
by the non-radiative term $\gamma_{nr}$ and is moderately affected by a change in $\gamma_r$. The increase in $Y$ that we observe is therefore essentially due to an increase of the radiative rate $\gamma_r$ 
of the QD because of the coupling to the NA.

\section{Conclusion}
In conclusion, we demonstrated deterministic coupling between single
semiconducting nanowire quantum dots and plasmonic nanoantennas using CL 
and electron-beam lithography with a precision of 
\unit{50}{\nano\meter}. Our method has the advantage of
relying only on the luminescence of the emitters for their precise localization.
It also grants full characterization of individual nanoemitters and antennas. CL
spectroscopy and LDOS imaging of individual NAs is demonstrated as a powerful
technique to experimentally determine antennas parameters for the fabrication of
coupled plasmonic-semiconductor emitters. Furthermore, we demonstrate two effects 
of the NA on the QD: an absorption enhancement of a factor 2, and a light emission 
enhancement due to radiative coupling to the antenna up to a factor 2.5 
in a region of high plasmonic losses,
extending the control of light emission from semiconducting nanostructures
towards the visible spectral region. The effect could be greatly increased for a
smaller QD-antenna distance, i.e. using a thinner NW shell. Implementation of
the method is a crucial step towards fabricating more complex and versatile
coupled structures. It can be applied to all kind of nanoemitters, aiming at
controlling their optical properties like their polarization response
\cite{Kukushkin2014,Casadei2015} or emission diagram
\cite{Curto2010,Belacel2013}.

\ack
We acknowledge the help of Institut N\'eel technical support teams Nanofab
(clean room) and  optical engineering (CL, Fabrice Donatini). This work
was supported by the French National Research Agency (project Magwires,
ANR-11-BS10-013 and Labex LANEF du Programme d'Investissements d'Avenir
ANR-10-LABX-51-01)

\section*{References}

\begin{thebibliography}{10}
	\expandafter\ifx\csname url\endcsname\relax
	\def\url#1{{\tt #1}}\fi
	\expandafter\ifx\csname urlprefix\endcsname\relax\def\urlprefix{URL }\fi
	\providecommand{\eprint}[2][]{\url{#2}}
	
	\bibitem{Birowosuto2014}
	Birowosuto M~D, Yokoo A, Zhang G, Tateno K, Kuramochi E, Taniyama H, Takiguchi
	M and Notomi M 2014 {\em Nat Mater\/} {\bf 13} 279–285 ISSN 1476-4660
	\urlprefix\url{http://dx.doi.org/10.1038/nmat3873}
	
	\bibitem{Ozel2013}
	Ozel T, Bourret G~R, Schmucker A~L, Brown K~A and Mirkin C~A 2013 {\em Advanced
		Materials\/} {\bf 25} 4515–4520 ISSN 0935-9648
	\urlprefix\url{http://dx.doi.org/10.1002/adma.201301367}
	
	\bibitem{Casadei2014}
	Casadei A, Pecora E~F, Trevino J, Forestiere C, Rüffer D, Russo-Averchi E,
	Matteini F, Tutuncuoglu G, Heiss M, Fontcuberta~i Morral A and Dal~Negro L
	2014 {\em Nano Lett.\/} {\bf 14} 2271--2278
	
	\bibitem{Casadei2015}
	Casadei A, Llado E~A, Amaduzzi F, Russo-Averchi E, Rüffer D, Heiss M, Negro
	L~D and Morral A~F~i 2015 {\em Scientific Reports\/} {\bf 5} 7651 ISSN
	2045-2322 \urlprefix\url{http://dx.doi.org/10.1038/srep07651}
	
	\bibitem{Ramezani2015}
	Ramezani M, Casadei A, Grzela G, Matteini F, Tütüncüoglu G, Rüffer D,
	Fontcuberta~i Morral A and Gómez~Rivas J 2015 {\em Nano Lett.\/} {\bf 15}
	4889–4895 ISSN 1530-6992
	\urlprefix\url{http://dx.doi.org/10.1021/acs.nanolett.5b00565}
	
	\bibitem{Curto2010}
	Curto A~G, Volpe G, Taminiau T~H, Kreuzer M~P, Quidant R and van Hulst N~F 2010
	{\em Science\/} {\bf 329} 930–933 ISSN 1095-9203
	\urlprefix\url{http://dx.doi.org/10.1126/science.1191922}
	
	\bibitem{Pfeiffer2010}
	Pfeiffer M, Lindfors K, Wolpert C, Atkinson P, Benyoucef M, Rastelli A, Schmidt
	O~G, Giessen H and Lippitz M 2010 {\em Nano Lett.\/} {\bf 10} 4555–4558
	ISSN 1530-6992 \urlprefix\url{http://dx.doi.org/10.1021/nl102548t}
	
	\bibitem{Nogues2013}
	Nogues G, Merotto Q, Bachelier G, Hye~Lee E and Dong~Song J 2013 {\em Applied
		Physics Letters\/} {\bf 102} 231112 ISSN 0003-6951
	\urlprefix\url{http://dx.doi.org/10.1063/1.4809831}
	
	\bibitem{Belacel2013}
	Belacel C, Habert B, Bigourdan F, Marquier F, Hugonin J~P, Michaelis~de
	Vasconcellos S, Lafosse X, Coolen L, Schwob C, Javaux C and et~al 2013 {\em
		Nano Lett.\/} {\bf 13} 1516–1521 ISSN 1530-6992
	\urlprefix\url{http://dx.doi.org/10.1021/nl3046602}
	
	\bibitem{Kukushkin2014}
	Kukushkin V~I, Mukhametzhanov I~M, Kukushkin I~V, Kulakovskii V~D, Sedova I~V,
	Sorokin S~V, Toropov A~A, Ivanov S~V and Sobolev A~S 2014 {\em Physical
		Review B\/} {\bf 90} ISSN 1550-235X
	\urlprefix\url{http://dx.doi.org/10.1103/PhysRevB.90.235313}
	
	\bibitem{Curto2013}
	Curto A~G, Taminiau T~H, Volpe G, Kreuzer M~P, Quidant R and van Hulst N~F 2013
	{\em Nat Comms\/} {\bf 4} 1750 ISSN 2041-1723
	\urlprefix\url{http://dx.doi.org/10.1038/ncomms2769}
	
	\bibitem{Hoang2016}
	Hoang T~B, Akselrod G~M and Mikkelsen M~H 2016 {\em Nano Lett.\/} {\bf 16}
	270–275 ISSN 1530-6992
	\urlprefix\url{http://dx.doi.org/10.1021/acs.nanolett.5b03724}
	
	\bibitem{Claudon2009}
	Claudon J, Bleuse J, Malik N~S, Bazin M, Jaffrennou P, Gregersen N, Sauvan C,
	Lalanne P and Gérard J~M 2009 {\em Nature Photon\/} {\bf 3} 116–116 ISSN
	1749-4893 \urlprefix\url{http://dx.doi.org/10.1038/nphoton.2009.287}
	
	\bibitem{Munsch2012}
	Munsch M, Claudon J, Bleuse J, Malik N~S, Dupuy E, Gérard J~M, Chen Y,
	Gregersen N and Mørk J 2012 {\em Physical Review Letters\/} {\bf 108} ISSN
	1079-7114 \urlprefix\url{http://dx.doi.org/10.1103/PhysRevLett.108.077405}
	
	\bibitem{Cremel2014}
	Cremel T, Elouneg-Jamroz M, Bellet-Amalric E, Cagnon L, Tatarenko S and Kheng K
	2014 {\em Phys. Status Solidi C\/} {\bf 11} 1263–1266 ISSN 1862-6351
	\urlprefix\url{http://dx.doi.org/10.1002/pssc.201300737}
	
	\bibitem{Artioli2013}
	Artioli A A, Rueda-Fonseca P P, Stepanov P P, Bellet-Amalric E, Den~Hertog M,
	Bougerol C, Genuist Y, Donatini F, André R R, Nogues G and et~al 2013 {\em
		Applied Physics Letters\/} {\bf 103} 222106 ISSN 0003-6951
	\urlprefix\url{http://dx.doi.org/10.1063/1.4832055}
	
	\bibitem{Rueda-Fonseca2014}
	Rueda-Fonseca P, Bellet-Amalric E, Vigliaturo R, den Hertog M, Genuist Y,
	André R, Robin E, Artioli A, Stepanov P, Ferrand D and et~al 2014 {\em Nano
		Lett.\/} {\bf 14} 1877–1883 ISSN 1530-6992
	\urlprefix\url{http://dx.doi.org/10.1021/nl4046476}
	
	\bibitem{Horodysky2005}
	Horodyský P and Hlídek P 2005 {\em physica status solidi (b)\/} {\bf 243}
	494–501 ISSN 0370-1972
	\urlprefix\url{http://dx.doi.org/10.1002/pssb.200541402}
	
	\bibitem{Ferrand2014}
	Ferrand D and Cibert J 2014 {\em Eur. Phys. J. Appl. Phys.\/} {\bf 67} 30403
	ISSN 1286-0050 \urlprefix\url{http://dx.doi.org/10.1051/epjap/2014140156}
	
	\bibitem{stepanov:tel-00994939}
	Stepanov P~S 2013 {\em Magneto-optical spectroscopy of magnetic semiconductor
		nanostructures\/} https://tel.archives-ouvertes.fr/tel-00994939
	Universit{\'e} de Grenoble
	\urlprefix\url{https://tel.archives-ouvertes.fr/tel-00994939}
	
	\bibitem{man2015}
	Man M~T and Lee H~S 2015 {\em Scientific Reports\/} {\bf 5} 8267 ISSN 2045-2322
	\urlprefix\url{http://dx.doi.org/10.1038/srep08267}
	
	\bibitem{Zielinski2013}
	Zieli\'nski M 2013 {\em Phys. Rev. B\/} {\bf 88} ISSN 1550-235X
	\urlprefix\url{http://dx.doi.org/10.1103/PhysRevB.88.115424}
	
	\bibitem{GarciadeAbajo2010}
	García~de Abajo F~J 2010 {\em Rev. Mod. Phys.\/} {\bf 82} 209–275 ISSN
	1539-0756 \urlprefix\url{http://dx.doi.org/10.1103/RevModPhys.82.209}
	
	\bibitem{Vesseur2007}
	Vesseur E~J~R, de~Waele R, Kuttge M and Polman A 2007 {\em Nano Lett.\/} {\bf
		7} 2843–2846 ISSN 1530-6992
	\urlprefix\url{http://dx.doi.org/10.1021/nl071480w}
	
	\bibitem{Supplementary}
	{See supplemental material about the main resonance peak in the antenna
		spectrum with Al$_2$O$_3$ thickness, Fourier plane imaging of NWQD
		fluorescence, lifetime measurements and cathodoluminescence of antennas at
		the excitation laser wavelength.}
	
	\bibitem{Davis2009}
	Davis T 2009 {\em Optics Communications\/} {\bf 282} 135–140 ISSN 0030-4018
	\urlprefix\url{http://dx.doi.org/10.1016/j.optcom.2008.09.043}
	
	\bibitem{Filter2012}
	Filter R, Qi J, Rockstuhl C and Lederer F 2012 {\em Physical Review B\/} {\bf
		85} ISSN 1550-235X
	\urlprefix\url{http://dx.doi.org/10.1103/PhysRevB.85.125429}
	
	\bibitem{Anger2006}
	Anger P, Bharadwaj P and Novotny L 2006 {\em Physical Review Letters\/} {\bf
		96} ISSN 1079-7114
	\urlprefix\url{http://dx.doi.org/10.1103/PhysRevLett.96.113002}
	
	\bibitem{Grzela2012}
	Grzela G, Paniagua-Domínguez R, Barten T, Fontana Y, Sánchez-Gil J~A and
	Gómez~Rivas J 2012 {\em Nano Lett.\/} {\bf 12} 5481–5486 ISSN 1530-6992
	\urlprefix\url{http://dx.doi.org/10.1021/nl301907f}
	
\end{thebibliography}
\providecommand{\newblock}{}

\end{document}